\documentclass[10pt,letterpaper,twocolumn]{article}

\usepackage[utf8]{inputenc}
\usepackage[T1]{fontenc}
\usepackage[top=16mm,bottom=18mm,left=16mm,right=16mm,columnsep=5mm]{geometry}
\usepackage{mathptmx}
\usepackage{authorpreprint}
\usepackage{amsmath}
\usepackage{amssymb}
\DeclareMathAlphabet{\mathcal}{OMS}{cmsy}{m}{n}
\SetMathAlphabet{\mathcal}{bold}{OMS}{cmsy}{b}{n}
\usepackage{graphicx}
\usepackage{float}
\usepackage{enumitem}
\usepackage{microtype}
\usepackage[font=footnotesize,labelfont=bf,labelsep=period]{caption}
\usepackage[numbers,sort&compress]{natbib}
\usepackage{xurl}
\usepackage{hyperref}

\floatstyle{ruled}
\newfloat{algorithm}{tbp}{loa}
\floatname{algorithm}{Algorithm}
\hypersetup{
  hidelinks,
  pdftitle={Physics-Based Differentiable Beamforming for Deep Learning-Based Contrast Enhanced Ultrasound Super-Resolution Imaging},
  pdfauthor={Anass Hameddine, Bram de Wilde, Michel Versluis, Jelmer M. Wolterink, Guillaume Lajoinie}
}

\title{Physics-Based Differentiable Beamforming for\\
Deep Learning-Based Contrast Enhanced\\
Ultrasound Super-Resolution Imaging}

\author{Anass Hameddine\textsuperscript{1}, Bram de Wilde\textsuperscript{1,2}, Michel Versluis\textsuperscript{1},\\[2pt]
Jelmer M. Wolterink\textsuperscript{2}, and Guillaume Lajoinie\textsuperscript{1}}
\affiliations{%
\textsuperscript{1}Physics of Fluids Group, Technical Medical (TechMed) Centre, Faculty of Science and Technology,\\
University of Twente, Enschede, The Netherlands\\[2pt]
\textsuperscript{2}Department of Applied Mathematics, Cardiovascular Health Technology Centre, Technical Medical (TechMed) Centre,\\
University of Twente, Enschede, The Netherlands}
\contactemails{\href{mailto:a.hameddine@utwente.nl}{a.hameddine@utwente.nl};
\href{mailto:g.p.r.lajoinie@utwente.nl}{g.p.r.lajoinie@utwente.nl}}
\date{}

\begin{document}

\maketitle

\begin{abstract}
Ultrasound is widely used for vascular imaging because it is non-invasive, real-time, and cost-effective. However, its spatial resolution remains limited by diffraction to an order of the acoustic wavelength. This limitation is particularly restrictive for imaging of the microvasculature and high-resolution vector-flow-imaging. In this work, we present a multi-channel physics-based deep learning approach to super-resolve monodisperse microbubbles directly from multi-channel radio-frequency (RF) data. We also present a novel image-based metric for model evaluation. The proposed pipeline combines a learnable deconvolution network with a differentiable beamforming layer, enabling a direct deconvolution of the microbubble (MB) signature prior to image formation. By operating on multi-channel RF data, the method aims to improve localization robustness in the presence of dense microbubble populations and deconvolve the lateral direction in a relatively low-frequency regime (1.7~MHz) where MBs exhibit nonlinear behavior. The model is trained and evaluated on simulated data using monodisperse lipid-coated MBs over a broad range of acoustic pressures and microbubble densities. The proposed method improves the signal-to-noise ratio (SNR) compared to the single-channel approach. It achieves sub-wavelength axial and lateral resolution while slightly improving robustness to dense microbubble populations outside the training domain. It also demonstrates generalizability across transmit configurations without requiring explicit training on each transmit configuration.
\end{abstract}

\begingroup\normalfont\fontsize{9}{11}\selectfont
\par
\endgroup

\section{Introduction}

\begin{figure*}[tbp]
  \centering
  \includegraphics[width=1\linewidth]{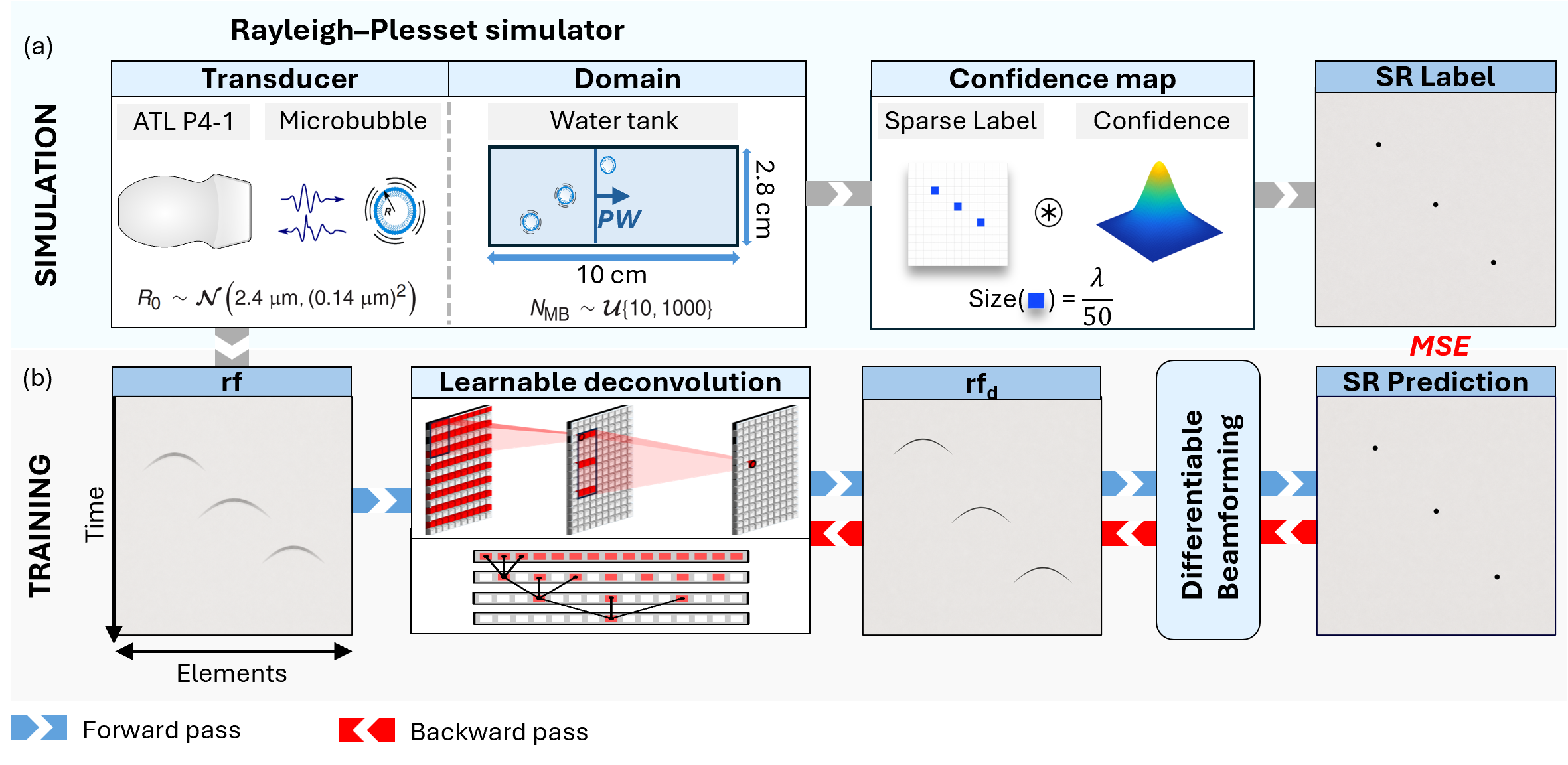}
  \caption{Overview of the proposed framework. (a) MB locations and their nonlinear acoustic responses generate simulated RF channel data. The corresponding binary GT map is convolved with a Gaussian kernel to produce a confidence map used as the super-resolution (SR) label, forming RF–label training pairs. (b) A learnable 2D deconvolution network processes the RF data, which are reconstructed by a differentiable beamforming layer. The MSE between the predicted and GT SR maps is backpropagated through the beamformer to optimize the deconvolution network.}
  \label{fig:sr_pipline} % Label for referencing the figure
\end{figure*}

Microvascular blood flow reflects tissue perfusion, vascular health, tumor angiogenesis, and disease progression, making it important in cardiovascular imaging, oncology, and neurology~\cite{Aziz2022-vb,Warboys2011-ub,Deffieux2021Functional}. Ultrasound is well suited to vascular imaging because it is non-invasive, real-time, portable, and cost-effective~\cite{szabo2013diagnostic}. However, diffraction limits its ability to resolve flow structures below the scale of the acoustic wavelength~\cite{8396283}.

Ultrasound localization microscopy (ULM) overcomes this limit by adapting localization principles from optical microscopy, such as PALM and FPALM~\cite{PALM1127344,FPALM}. ULM treats contrast microbubbles (MBs) as isolated acoustic point scatterers and localizes them across successive ultrafast frames to reconstruct vascular networks at sub-wavelength resolution~\cite{Christensen-Jeffries2020-in}.

Conventional ULM nevertheless requires sparse MB distributions to prevent overlapping point spread functions and speckle interference. Consequently, few MBs can be localized per frame, requiring many frames and acquisition times of typically tens of minutes~\cite{Christensen-Jeffries2020-in}. 

Deep learning has improved MB detection and localization while reducing computational cost~\cite{9257449,9251561,Luan_2023}. Most approaches operate on beamformed images~\cite{10130654,9715135,9062600}, while recent methods directly process radio-frequency (RF) data to exploit the information contained in the received signals~\cite{10506070,9755198,10858770,9130724,9715054}.

We previously introduced a single-channel RF-based method that deconvolves one-dimensional signals from the center transducer element~\cite{9755198}. Because each element receives a delayed version of the same MB signature, the model can be trained on one element and applied channel-wise during inference. Although computationally lightweight, this strategy does not exploit inter-channel spatial coherence, limiting lateral deconvolution and MB separation at high densities, where substantial signal overlap can occur. In our previous study, an overlap rate of up to $95\%$ was observed for 1000 MBs~\cite{9755198}.

Here, we propose a physics-based multi-channel framework that super-resolves MB signatures directly from two-dimensional RF data. A learnable deconvolution network is combined with an embedded differentiable beamformer, enabling the model to exploit inter-channel information within a physically interpretable and data-efficient formulation. The framework is intended for integration with vector flow imaging (VFI) to support high-resolution velocity estimation in larger vessels under dense MB conditions~\cite{2021210454,doi:10.1148/radiol.2018172979}, and can also be interfaced with ULM to enable localization and tracking of MB centroids under dense conditions.

The models are trained on simulated monodisperse lipid-coated MB responses spanning acoustic pressures of $5$--$250~\mathrm{kPa}$ and populations of $10$--$1000$ MBs per sample. Simulations are performed at $1.7$~MHz, a low-frequency regime relevant to deep imaging and nonlinear MB behavior~\cite{SEGERS20181482,OVERVELDE20102080,9755198}. The contributions are threefold. First, we extend the single-channel RF deconvolution approach to multi-channel data to improve lateral resolution and handle dense MB conditions. Second, we incorporate reconstruction physics through a tensor-based differentiable beamformer, resulting in a data-efficient gray-box model with improved generalization across transmit configurations. Third, we introduce an image-based evaluation pipeline that jointly assesses MB detection, localization error, and localization uncertainty.

\section{Materials and Methods}
\label{sec:methods}

\subsection{Ultrasound data simulation}

Training and validation data were generated using the RF simulator developed by Blanken et al.~\cite{9755198}. Each sample contained monodisperse MBs with radii \(R_0 \sim \mathcal{N}\left(2.4~\mu\mathrm{m},(0.14~\mu\mathrm{m})^2\right)\), randomly distributed in a two-dimensional aqueous domain with a sound speed of \(c=1480~\mathrm{m/s}\). The domain measured \(2.8~\mathrm{cm}\) in width and \(10~\mathrm{cm}\) in depth, with MB depths restricted to \(0.37\leq z\leq9.63~\mathrm{cm}\). The upper bound ensured complete recording of echoes near the domain boundary, while the lower bound avoided excessive near-surface amplification from the \(1/r_{\mathrm{rx}}\) receive-distance factor, where \(r_{\mathrm{rx}}\) is the bubble-to-element distance.

For each sample, the number of MBs, \(N_{\mathrm{mb}}\), was uniformly sampled between $10$ and $1000$, spanning a broad range of densities and PSF-overlap levels. Additional out-of-distribution test samples containing up to \(10{,}000\) MBs were generated to evaluate robustness under denser conditions. The simulated transducer was based on the P4-1 phased array (Philips ATL), operating at \(1.7~\mathrm{MHz}\) with a one-cycle transmit pulse. The peak acoustic pressure was sampled as \(P_{\mathrm{ac,max}}\sim\mathcal{U}(5~\mathrm{kPa},250~\mathrm{kPa})\).

Two extensions were added to the original simulator~\cite{9755198} for evaluation purposes, as summarized in Table~\ref{tab:tx_config}. First, transmit steering was modeled by replacing the plane-wave (PW) transmit propagation time \(\tau_{\mathrm{tx}}(x,z)=z/c\) with
\begin{equation}
    \tau_{\mathrm{tx}}(x,z,\theta) = \frac{z\cos\theta+x\sin\theta}{c}
\end{equation}
Second, diverging-wave (DW) transmission was modeled using a virtual source at \(x_v=0\) and \(z_v=-W/(2\tan(\alpha))\), where \(W\) is the aperture width and \(\alpha\) is the wavefront half-angle~\cite{PERROT2021106309}. Because training used unsteered PW transmissions, the steered PW transmissions and the different DW geometry provided additional out-of-distribution test scenarios. Geometric spreading was modeled using an amplitude decay proportional to \(1/\sqrt{r_v}\), where \(r_v\) is the distance from the virtual source.

\begin{table}[tbp] %[H]
\centering
\caption{Simulated data for training and evaluation}
\label{tab:tx_config}
\renewcommand{\arraystretch}{1.15}
\begin{tabular}{ccc}
    \hline
    \shortstack{\textbf{Transmission}\\\textbf{Type}} & \shortstack{\textbf{Steering}\\\textbf{Angle ($^\circ$)}} & \textbf{Usage} \\
    \hline
     PW & $0^\circ$ & Training \\
     PW & $0^\circ : 1^\circ : 20^\circ$ & Evaluation \\
     DW & $0^\circ$ & Evaluation \\
    \hline
\end{tabular}
\end{table}

\subsection{Super-resolution Model}

Consider RF channel data \(\mathbf{rf}\in\mathbb{R}^{N_e\times N_t}\) generated by an MB distribution \(\mathbf{mb}\in\mathbb{R}^{N_{mb}\times 2}\), where each row of \(\mathbf{mb}\) contains the lateral and axial coordinates of one MB. Here, \(N_e\), \(N_t\), and \(N_{mb}\) denote the numbers of active receive elements, temporal samples, and MBs, respectively. We seek a nonlinear mapping \(\Phi:\mathbb{R}^{N_e\times N_t}\rightarrow\mathbb{R}^{N_x\times N_z}\) that estimates a super-resolved MB response map, where \(N_x\) and \(N_z\) denote the numbers of lateral and axial image pixels.

We decompose \(\Phi\) into a learnable RF deconvolution function \(d_{\Theta}\), followed by a beamforming function \(b_f\):
\begin{equation}
\Phi(\mathbf{rf})=(b_f\circ d_{\Theta})(\mathbf{rf}).
\label{eq:non-linear-mapping}
\end{equation}
Here \(d_{\Theta}:\mathbb{R}^{N_e\times N_t}\rightarrow\mathbb{R}^{N_e\times N_t}\) is parameterized by the trainable parameters \(\Theta\), and \(b_f:\mathbb{R}^{N_e\times N_t}\rightarrow\mathbb{R}^{N_x\times N_z}\) maps the deconvolved RF data to the reconstructed image.

\subsection{Learnable Deconvolution Model}

The deconvolution network \(d_{\Theta}\) maps the multi-channel RF data \(\mathbf{rf}\in\mathbb{R}^{N_e\times N_t}\) to deconvolved RF data \(\mathbf{rf}_d\) before differentiable beamforming. The input and output therefore have the same dimensions, \(N_e\times N_t\).

The architecture is adapted from the single-channel RF deconvolution model introduced in~\cite{9755198}, which processes each RF channel independently using a one-dimensional dilated CNN. Because an MB response can extend over approximately \(2~\mu\mathrm{s}\), corresponding to about \(125\) temporal samples, the network requires a large temporal receptive field to capture the complete response and its depth-dependent context, including responses from neighboring MBs. The original model therefore uses exponentially increasing dilation factors \(d_m=2^m\), with \(m=0,\ldots,L-1\). For a kernel size of \(3\) and \(L\) convolutional layers, the receptive field is \(N_r=2^{L+1}-1\), compared with \(N_r=2L+1\) for an equivalent non-dilated CNN.

We retain this temporal formulation and extend the convolutions along the receive-element dimension using anisotropic two-dimensional kernels. The temporal kernel size is fixed at \(k_t=3\) with exponentially increasing dilation, whereas the lateral kernel size \(k_x\) controls the extent across adjacent receive channels using unit dilation. Setting \(k_x=1\) recovers the original independent single-channel formulation, whereas \(k_x>1\) introduces local lateral coupling and exploits inter-channel spatial coherence before beamforming. We use \(k_x=7\), allowing each layer to combine information from a given channel and up to six neighboring receive channels.

The network is fully convolutional. The first convolution maps the RF input to \(64\) feature channels, which are maintained throughout the intermediate dilated layers. Each hidden convolution is followed by batch normalization and a rectified linear unit (ReLU). A final linear \(1\times1\) convolution combines the \(64\) feature channels into the deconvolved RF output \(\mathbf{rf}_d\). Reflection padding along the temporal dimension and zero-padding along the receive-element dimension preserve the input dimensions, such that \(\mathbf{rf}_d\) and \(\mathbf{rf}\) have the same size.

\subsection{Beamforming Layer}

As defined in~\eqref{eq:non-linear-mapping}, the beamforming operator \(b_f\) is implemented as a differentiable layer following the deconvolution network \(d_{\Theta}\). During end-to-end training, \(d_{\Theta}\) maps the multi-channel RF input to a deconvolved representation \(\mathbf{rf}_d\), which \(b_f\) reconstructs into an image. This formulation enables image-domain supervision while preserving the physical structure of the ultrasound reconstruction pipeline.

We implement and compare two differentiable beamforming operators: Stolt's \(f\)-\(k\) migration~\cite{6587395} and Delay-and-Sum (DAS). Stolt's \(f\)-\(k\) migration operates in the frequency--wavenumber domain and focuses each RF diffraction parabola at its apex through spectral remapping. In contrast, DAS operates in the time domain and reconstructs the image through delay-based coherent summation across receive elements.

\subsubsection{\texorpdfstring{Stolt's $f$-$k$ migration}{Stolt's f-k migration}}

Originally developed for seismic imaging, Stolt's \(f\)-\(k\) migration reconstructs the scatterer distribution by time-reversing wave propagation in the Fourier domain under the exploding reflector model (ERM). The ERM treats scatterers as simultaneous sources at \(t=0\) that emit upward-propagating waves, enabling efficient reconstruction using the fast Fourier transform (FFT) and frequency-domain interpolation~\cite{Stolt1978-vg}.

For PW ultrasound imaging, the ERM was adapted to account for two-way propagation, particularly near the RF parabola apex, where most of the energy is concentrated~\cite{6587395}. We further correct the pulse-induced axial shift of the beamformed image by multiplying the spectrum by \(e^{-i2\pi f\tau^{(\mathrm{p})}}\), where \(\tau^{(\mathrm{p})}\) is computed as

\begin{equation}
    \tau^{(\mathrm{p})} = \frac{\operatorname*{argmax}_{i} \left|\mathcal{H}({p})\right|}{f_s^{(\mathrm{p})}}.
\label{eq:pulse_dur_corr}
\end{equation}

Here, \(p\), \(\mathcal{H}\{\cdot\}\), and \(f_s^{(\mathrm{p})}\) denote the pulse, Hilbert transform, and pulse sampling frequency, respectively.

\subsubsection{Delay-And-Sum}

\begin{figure}[tbp]
    \centering
    \includegraphics[width=\linewidth]{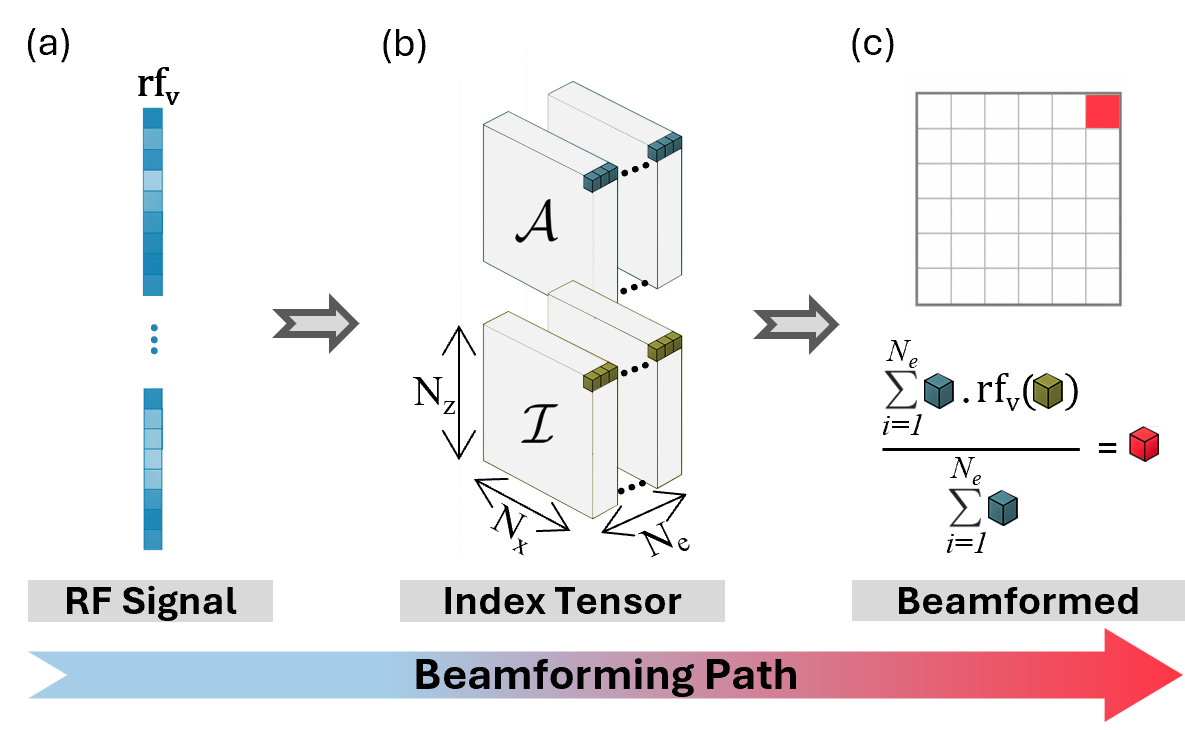}
    \caption{Tensor-based DAS beamforming. (a) The multi-channel RF data are vectorized as \(\mathbf{rf}_{\mathrm{v}}\). (b) The tensors \(\mathcal{I}\) and \(\mathcal{A}\) store the delay indices and apodization weights, respectively. (c) RF samples gathered using \(\mathcal{I}\) are weighted by \(\mathcal{A}\), summed, and normalized by the number of active elements to form the beamformed pixel at \((x,z)\), shown in red.}
    \label{fig:das}
\end{figure}

DAS beamforming reconstructs an image by coherently summing the delayed RF signals across transducer elements. For integration into the deep learning framework, we reformulate DAS using tensor operations.

Let \(\mathbf{rf}\in\mathbb{R}^{N_e\times N_t}\) denote the RF data and let the reconstructed image contain \(N_x\times N_z\) pixels. The delay-index and apodization tensors are denoted by \(\mathcal{I}\in\mathbb{N}^{N_x\times N_z\times N_e}\) and \(\mathcal{A}\in\{0,1\}^{N_x\times N_z\times N_e}\), respectively. For each image point \((x,z)\) and receive element \(e\), \(\mathcal{I}(x,z,e)\) stores the RF sample index associated with the propagation delay, whereas \(\mathcal{A}(x,z,e)\) stores the receive-apodization weight, as illustrated in Fig.~\ref{fig:das}.

The total propagation time is
\begin{equation}
t(x,z,e)
=
t_{\mathrm{tx}}(x,z,\theta)
+ t_{\mathrm{rx}}(x,z,e)
+ \tau^{(\mathrm{p})}.
\end{equation}
Here, \(t_{\mathrm{tx}}(x,z,\theta)\) is the transmit propagation time at steering angle \(\theta\), \(t_{\mathrm{rx}}(x,z,e)\) is the receive propagation time between \((x,z)\) and element \(e\), and \(\tau^{(\mathrm{p})}\) is the time-to-peak correction defined in~\eqref{eq:pulse_dur_corr}. The corresponding sample index is stored as \(\mathcal{I}(x,z,e)=\left\lfloor F_s t(x,z,e)\right\rceil\), where \(F_s\) is the RF sampling frequency and \(\left\lfloor\cdot\right\rceil\) denotes nearest-integer rounding.

A binary angular apodization limits the receive aperture. The angle of view between \((x,z)\) and the receive element at \((x_e,0)\) is \(\alpha=\arctan\left(|x_e-x|/z\right)\). The corresponding weight is set to \(\mathcal{A}(x,z,e)=1\) when \(\alpha\leq\alpha_{\mathrm{th}}\) and to \(0\) otherwise, with \(\alpha_{\mathrm{th}}=\pi/4\).

Using \(\mathcal I\), the delayed samples are gathered from \(\mathbf{rf}_{\mathrm v}=\operatorname{vec}(\mathbf{rf})\), and the beamformed RF signal is computed as
\begin{equation}
    B_{\mathrm{RF}}(x,z)
    =
    \frac{
        \sum_{e=1}^{N_e}
        \mathcal{A}(x,z,e)\,
        \mathbf{rf}_{\mathrm{v}}\!\left(\mathcal{I}(x,z,e)\right)
    }{
        \sum_{e=1}^{N_e}\mathcal{A}(x,z,e)
    }.
\end{equation}
The final envelope image is \(B(x,z)=\left|\mathcal{H}\{B_{\mathrm{RF}}(x,z)\}\right|\).

To reduce memory usage, \(\mathcal I\) and \(\mathcal A\) are generated in subsets of \(C\) receive elements. The delayed samples from each subset are gathered and accumulated into the output, reducing the required tensor size from \(N_xN_zN_e\) to \(N_xN_zC\), corresponding to an approximate memory reduction factor of \(N_e/C\). This tensor-based DAS formulation avoids element-wise loops and provides a tunable memory footprint suitable for deep learning frameworks.

\subsection{Model Training}

The model was trained using a post-beamforming mean squared error (MSE) loss against Gaussian-smoothed target labels. The output grid contains \(1603\times5727\) pixels, whereas the densest simulated sample contains only \(1000\) MBs, corresponding to approximately \(99.99\%\) sparsity. Because such sparse binary targets provide limited spatial supervision, Gaussian smoothing was applied to produce a denser and more spatially tolerant target, facilitating optimization and reducing sensitivity to the selected pixel size.

The MB distribution was first represented as a binary image \(\mathbf{y}\), with MB locations set to \(1\) and all other pixels to \(0\). The smoothed target was then computed as \(\mathbf{y}_s=G_{\sigma_x,\sigma_z}\ast\mathbf{y}\) using an isotropic two-dimensional Gaussian kernel with \(\sigma_x=\sigma_z=\sigma\). Its full width at half maximum was set to \(3\lambda/50\), with \(\mathrm{FWHM}=2\sqrt{2\ln(2)}\,\sigma\), thereby enforcing a depth-independent target shape. The training loss between the beamformed prediction \(\hat{\mathbf{y}}\) and the smoothed target was
\begin{equation}
\mathcal{L}_{\mathrm{MSE}}(\mathbf{y}_s,\hat{\mathbf{y}})
=
\frac{1}{N}
\sum_{i=1}^{N}
\left(y_{s,i}-\hat{y}_i\right)^2.
\label{eq:loss}
\end{equation}
Here, \(N\) denotes the number of output pixels. The deconvolution network was optimized solely through the loss in~\eqref{eq:loss}, while being implicitly constrained by the beamforming physics and the prescribed lateral and axial target resolutions.

Following~\cite{9755198}, the dataset was divided into \(1024\), \(960\), and \(960\) samples for training, validation, and testing, respectively. The network was trained for \(1250\) epochs using Adam with a learning rate of \(10^{-3}\), no weight decay, and Xavier initialization. A batch size of \(2\) with gradient accumulation over four steps provided an effective batch size of \(8\).

Each RF sample \(\mathbf{rf}\in\mathbb{R}^{N_e\times N_t}\) contained \(N_e=96\) receive channels and \(N_t=8446\) temporal samples. The RF data were independently z-score normalized by subtracting the sample mean and dividing by its standard deviation. Training used model compilation and FP16 mixed precision on an NVIDIA Quadro RTX 6000 GPU with \(24\) GB of memory.

\subsection{Evaluation Metrics}

At the MB level, detection performance is quantified using the \(F_1\) score, defined as
\begin{equation}
    F_1
    =
    \frac{2\,\mathrm{Precision}\,\mathrm{Recall}}
    {\mathrm{Precision}+\mathrm{Recall}}
    =
    \frac{2\,\mathrm{TP}}
    {2\,\mathrm{TP}+\mathrm{FP}+\mathrm{FN}}.
\end{equation}
Here, TP, FP, and FN denote true-positive, false-positive, and false-negative detections, respectively.

Many ULM pipelines extract discrete MB coordinates using local-maximum detection~\cite{9130724}, difference-of-Gaussians detection~\cite{9715054}, connected components and center-of-mass localization~\cite{9062600}, or Gaussian fitting, matched filtering, normalized cross-correlation, morphological reconstruction, and weighted centroids~\cite{10497610}. Although effective, these methods reduce each response to a single coordinate and often require empirically selected thresholds, peak-selection rules, or PSF estimates. Coordinate-based evaluation therefore determines whether the estimated center lies within a prescribed tolerance but provides limited information about the spatial quality of the response.

\begin{figure}[tbp]
    \centering
    \includegraphics[width=\linewidth]{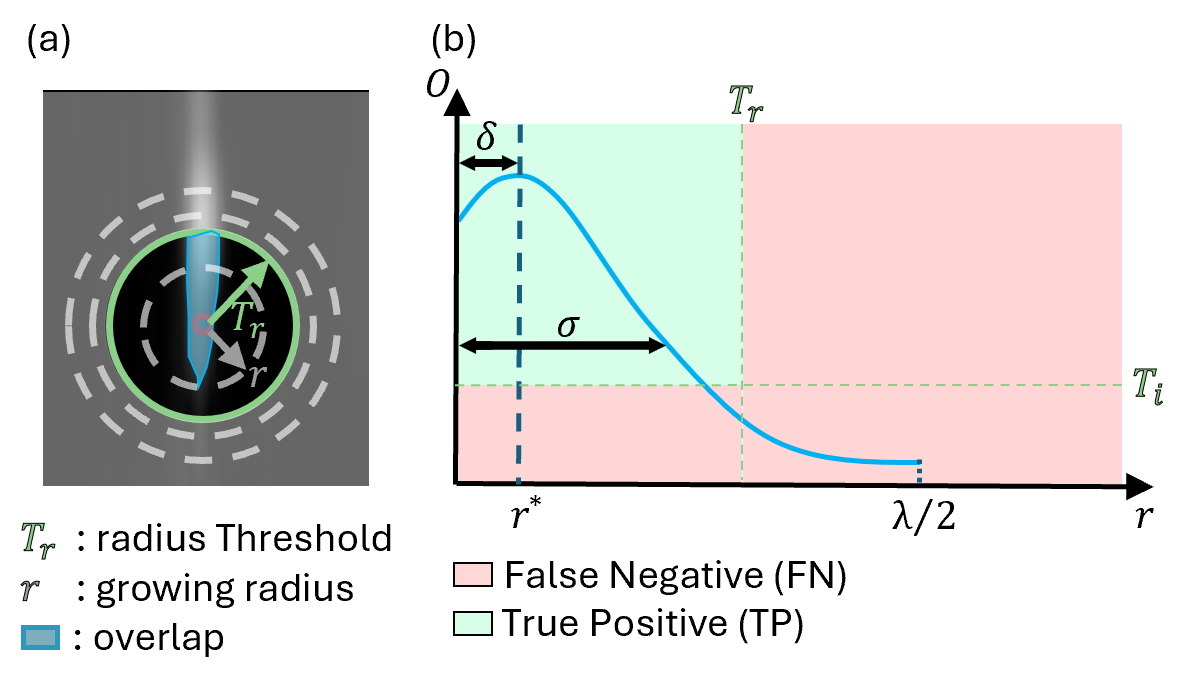}
    \caption{Illustration of the radial expansion metric (REM). (a) A radially expanding disk measures the overlap with the predicted response around the ground-truth MB location (blue), with \(T_r\) denoting the maximum accepted radius. (b) The optimal radius \(r^\ast\) identifies the MB peak intensity, with \(T_i\) shown for reference. The localization error \(\delta\) and uncertainty \(\sigma\) characterize the spatial accuracy and extent of the predicted response.}
    \label{fig:TPFN}
\end{figure}

To additionally characterize each MB’s local response, we introduce the radial expansion metric (REM), which measures the area-normalized predicted response within a radially expanding disk and jointly assesses detection, localization error, and response uncertainty. As illustrated in Fig.~\ref{fig:TPFN}, REM uses a radial tolerance \(T_r=\lambda/25\) and an intensity threshold \(T_i=1/e\). The former constrains the response around the GT location, whereas the latter reflects the Gaussian target and rejects weak or spatially inconsistent responses.

Algorithm~\ref{alg:tp_fp_fn} evaluates each GT position \(\mathbf{p}_i\) using disks \(\mathcal{D}_r(\mathbf{p}_i)\), with \(r\in[0,\lambda/2]\). The radius \(r_i^\ast\) maximizes the mean response \(O_i(r)\), and \(\mathbf{q}_i^\ast\) identifies the response maximum within the selected disk. A prediction is classified as a TP when \(\hat{\mathbf{y}}(\mathbf{q}_i^\ast)\geq T_i\) and \(r_i^\ast\leq T_r\), otherwise, it is classified as an FN. After removing the accepted TP prediction regions, connected components in the thresholded residual image with areas of at least \(A_{\min}=5\) pixels are counted as FPs.

\begin{figure*}[tbp]
    \centering
    \includegraphics[width=1\linewidth]{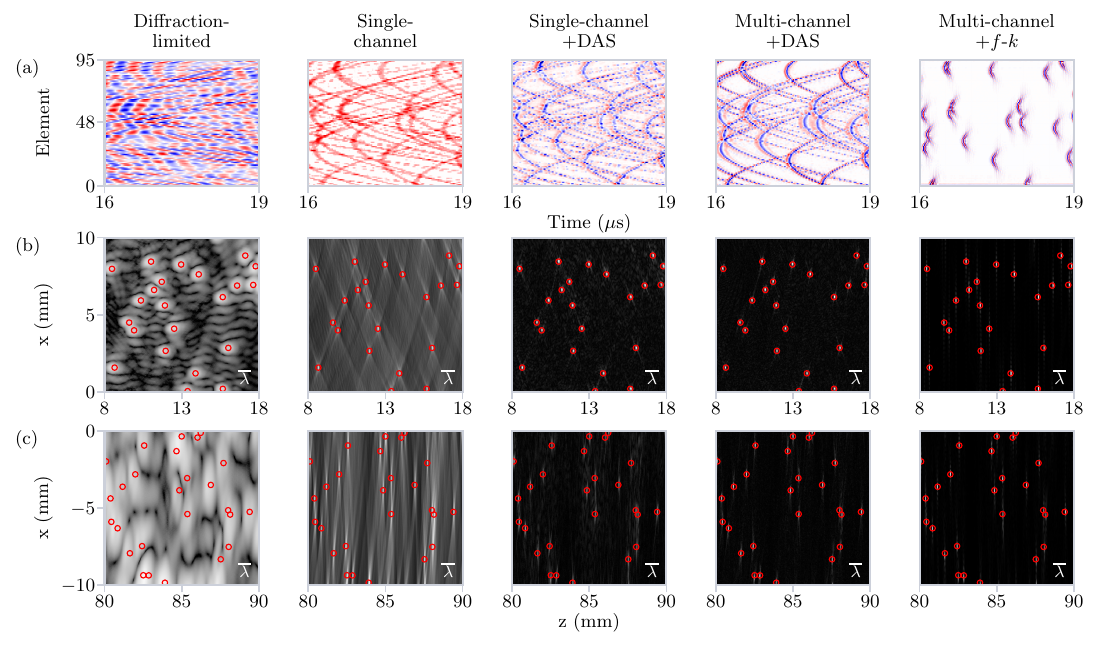}
    \caption{Representative results for a dense population of 988 MBs. The  red circles mark ground-truth (GT) locations. (a) Deconvolved multi-channel RF signals for each configuration. (b) Near-field reconstructions showing higher SNR for the multi-channel models than the Single-channel baseline. (c) Far-field reconstructions showing that explicit multi-channel processing improves lateral deconvolution and MB-response separation over the Single-channel model.}
    \label{fig:qualitative_res}
\end{figure*}

\begin{algorithm}[tbp]
    \caption{TP, FP, and FN identification using REM}
    \label{alg:tp_fp_fn}
\begin{enumerate}[label=\arabic*:,leftmargin=2.5em,itemsep=0.2em,parsep=0pt]
\item[] \textbf{Require:} GT positions \(\{\mathbf{p}_i\}_{i=1}^{N_{mb}}\), prediction \(\hat{\mathbf{y}}\), thresholds \(T_i,T_r\), minimum area \(A_{\min}\)
\item[] \textbf{Ensure:} \(TP,FP,FN\)
\item \(TP\gets0,\;FN\gets0,\;\mathbf{y}_{\mathrm{TP}}\gets\mathbf{0}\)
\item \textbf{for} \(i=1\) to \(N_{mb}\) \textbf{do}
    \item \hspace*{1em}Compute \(O_i(r)\gets
    \frac{\sum_{\mathbf{q}\in\mathcal{D}_r(\mathbf{p}_i)}
    \hat{\mathbf{y}}(\mathbf{q})}
    {|\mathcal{D}_r(\mathbf{p}_i)|}\),
    \(r\in[0,\lambda/2]\)
    \item \hspace*{1em}\(r_i^\ast\gets\operatorname*{arg\,max}_{r}O_i(r)\)
    \item \hspace*{1em}\(\mathbf{q}_i^\ast\gets
    \operatorname*{arg\,max}_{\mathbf{q}\in
    \mathcal{D}_{r_i^\ast}(\mathbf{p}_i)}
    \hat{\mathbf{y}}(\mathbf{q})\)
    \item \hspace*{1em}\textbf{if} \(\hat{\mathbf{y}}(\mathbf{q}_i^\ast)\geq T_i~
    \textbf{and}~r_i^\ast\leq T_r\) \textbf{then}
        \item \hspace*{2em}\(TP\gets TP+1\)
        \item \hspace*{2em}\(\mathbf{y}_{\mathrm{TP}}(\mathbf{q})
        \gets\hat{\mathbf{y}}(\mathbf{q})\),
        \(\forall\mathbf{q}\in\mathcal{D}_{r_i^\ast}(\mathbf{p}_i)\)
    \item \hspace*{1em}\textbf{else}
        \item \hspace*{2em}\(FN\gets FN+1\)
    \item \hspace*{1em}\textbf{end if}
\item \textbf{end for}
\item \(\mathbf{R}\gets\hat{\mathbf{y}}-\mathbf{y}_{\mathrm{TP}}\)
\item Threshold \(\mathbf{R}\) at \(T_i\) to obtain \(\mathbf{B}\)
\item Remove components in \(\mathbf{B}\) with area \(<A_{\min}\)
\item \(FP\gets\#\mathrm{CC}(\mathbf{B})\)
\end{enumerate}
\end{algorithm}

TPs, FPs, and FNs were identified after applying controlled depth-dependent intensity saturation to the beamformed images. This limits the dominance of strong near-field responses over weaker far-field MBs that remain well localized and distinguishable from the background. The saturation level was derived from the expected MB occupancy rather than selected empirically.

We used half the maximum training density, \(N_{\mathrm{ref}}=N_{mb,\max}/2=500\), where \(N_{mb,\max}=1000\). Each MB target was modeled as a Gaussian with \(\mathrm{FWHM}=3\) pixels, giving \(\sigma_{\mathrm{G}}=\mathrm{FWHM}/2.355\). Using an effective radius of \(2\sigma_{\mathrm{G}}\), the occupied area was approximated as \(A_{\mathrm{MB}}\approx\pi(2\sigma_{\mathrm{G}})^2\approx20\) pixels. The saturation percentage was then
\begin{equation}
    p_{\mathrm{sat}}
    =
    \frac{N_{\mathrm{ref}}A_{\mathrm{MB}}}
    {N\gamma}\times100\%.
\end{equation}
Here, \(N\) is the number of reconstructed pixels and \(\gamma=4\) restricts saturation to the high-intensity core rather than the full Coordinate-based evaluation quantifies detection performance and localization accuracy but does not directly characterize the spatial extent or intensity distribution of each predicted MB response. MB response. For \(N=1603\times5727=9{,}180{,}381\), this gives \(p_{\mathrm{sat}}\approx0.027\%\), corresponding to approximately \(2.5\times10^3\) pixels, or five high-intensity pixels per reference MB. Saturation therefore primarily affects the brightest MB cores.

Performance is reported using the \(F_1\) score, localization uncertainty \(\sigma\), and localization error \(\delta\). Curves represent means across evaluated samples, while shaded regions and error bars indicate one standard deviation.

\subsection{Models}

We evaluate five configurations spanning the previously published single-channel baseline to the fully two-dimensional model. The Single-channel model reimplements the original approach by Blanken et al.~\cite{9755198}. Single-channel+DAS applies the same network independently to all \(96\) RF lines, reconstructs their outputs using differentiable DAS, and computes the MSE loss in the image domain. This weakly couples the RF lines during training and enables assessment of the benefit of explicit two-dimensional RF processing. Multi-channel+DAS and Multi-channel+\(f\)-\(k\) combine the proposed two-dimensional network with differentiable DAS and \(f\)-\(k\) migration, respectively. Finally, Multi-channel+Id replaces the beamforming layer with an identity operator, requiring the network to learn RF deconvolution and image reconstruction jointly and thereby assessing the contribution of explicit beamforming physics.

\section{Results}
\label{sec:results}

\subsection{Impact of Multi-Channel Information and Beamforming Constraints}

\begin{figure}[tbp]
    \centering
    \includegraphics[width=\linewidth,height=0.63\textheight,keepaspectratio]{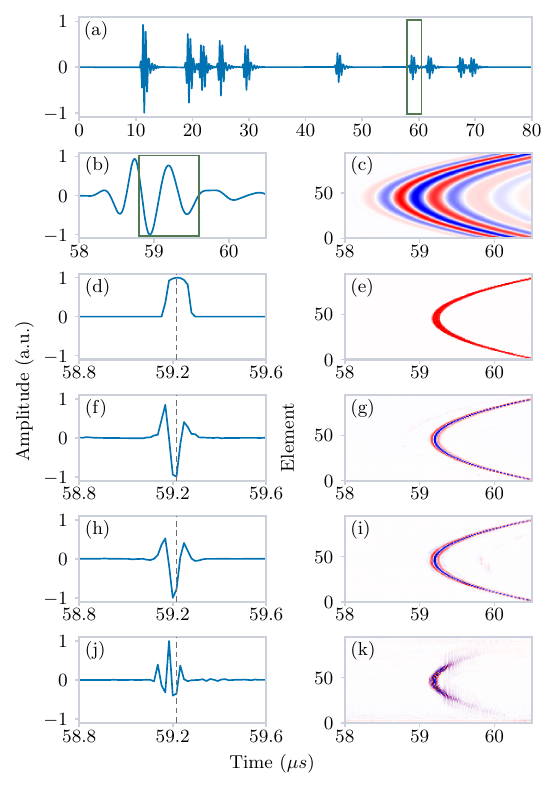}
    \caption{Qualitative comparison of RF deconvolution across methods. Columns show the central-element RF trace and the corresponding 2-D multi-channel data. (a–c) Diffraction-limited inputs before deconvolution. (d,e) Single-channel output, exhibiting an impulse-like response along the processed trace. (f,g) DAS-constrained single-channel output, yielding a tripolar waveform. (h,i) The corresponding multi-channel formulation, which retains a similar tripolar structure. (j,k) Stolt \(f\)-\(k\) migration-constrained output, exhibiting a sharper and more localized tripolar-like waveform.}
    \label{fig:rf}
\end{figure}

We evaluated the proposed method on simulated RF data using the image-based metrics in Algorithm~\ref{alg:tp_fp_fn}. Fig.~\ref{fig:qualitative_res} shows a representative dense case containing 988 MBs, located mainly near \(8.5\)~cm depth. Strongly overlapping diffraction-limited responses obscure individual MBs. The Single-channel model improves axial separation but remains laterally limited because it does not exploit inter-channel spatial coherence.

\begin{figure*}[tbp]
    \centering
    \includegraphics[width=\linewidth]{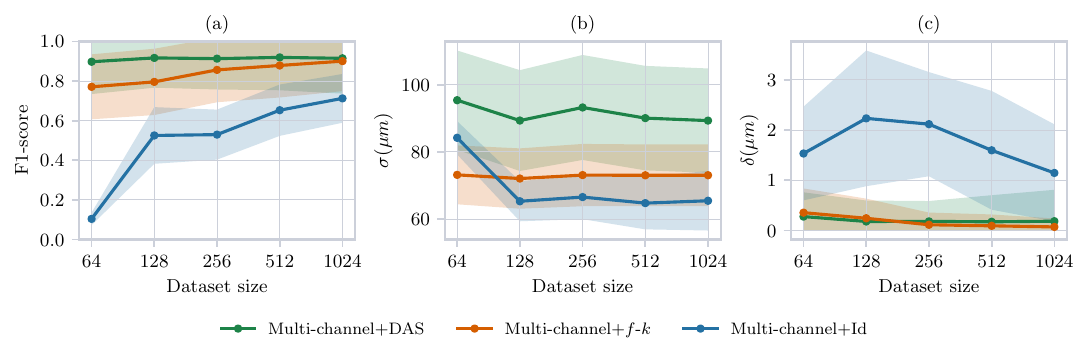}
    \caption{(a) $F_1$ score, (b) uncertainty, (c) and localization error as a function of the training dataset size. (b) and (c) are based on TP predictions. Physics-based models achieve higher detection performance with fewer training samples, indicating improved data efficiency compared with the fully data-driven baseline.}
    \label{fig:data_size}
\end{figure*}

Table~\ref{tab:metrics} quantifies these observations. Single-channel achieves an \(F_1\) score of \(0.27\), with localization uncertainty and error of \(252.13~\mu\mathrm{m}\) and \(1.30~\mu\mathrm{m}\), respectively. Its low \(F_1\) score reflects residual MB tails and background intensities above the \(1/e\) threshold, which are counted as FPs. Adding differentiable DAS increases the \(F_1\) score to \(0.90\) and reduces the error to \(0.38~\mu\mathrm{m}\). By coupling the RF lines through the image-domain reconstruction loss, Single-channel+DAS improves MB separation and background suppression.

The multi-channel models improve performance further. Multi-channel+DAS achieves the highest \(F_1\) score of \(0.92\), with uncertainty and error of \(89.32~\mu\mathrm{m}\) and \(0.19~\mu\mathrm{m}\), respectively. Multi-channel+\(f\)-\(k\) attains a comparable \(F_1\) score of \(0.90\) and the lowest error of \(0.07~\mu\mathrm{m}\), whereas Multi-channel+Id obtains \(0.71\) and \(1.15~\mu\mathrm{m}\), respectively, despite producing compact responses. These results show that multi-channel processing improves lateral separation and that reconstruction constraints benefit both detection and localization. Fig.~\ref{fig:data_size} further shows that the physics-constrained models retain strong performance with only 64 training samples.

The improvement is also evident in the deconvolved RF domain (Figs.~\ref{fig:qualitative_res} and~\ref{fig:rf}), where the multi-channel outputs exhibit lower background levels and more structured MB responses. Rather than collapsing to Dirac impulses, they retain beamformer-dependent waveforms. Multi-channel+DAS produces a higher-frequency tripolar pulse with amplitude distributed along the RF parabola, whereas Multi-channel+\(f\)-\(k\) produces a similar response with energy concentrated near the parabola apex. The differentiable beamformer therefore constrains the intermediate RF representation toward structured and physically interpretable responses.

\begin{table}[tbp]
\centering
\caption{Single-channel vs multi-channel performance.}
\label{tab:metrics}
\small
\setlength{\tabcolsep}{3pt}
\renewcommand{\arraystretch}{1.15}
\begin{tabular}{@{}lccc@{}}
\hline
\textbf{Model} &
\shortstack{\textbf{$F_1$}\\\textbf{score}} &
\shortstack{\textbf{Mean}\\\textbf{Uncertainty}\\\textbf{($\mu$m)}} &
\shortstack{\textbf{Mean Loc.}\\\textbf{error}\\\textbf{($\mu$m)}} \\
\hline
Single-channel         & $0.27$ & $252.13$ & $1.30$ \\
Single-channel+DAS     & $0.90$ & $118.66$ & $0.38$ \\
Multi-channel+DAS      & $0.92$ & $89.32$  & $0.19$ \\
Multi-channel+$f$-$k$  & $0.90$ & $73.07$  & $0.07$ \\
Multi-channel+Id       & $0.71$ & $65.47$  & $1.15$ \\
\hline
\end{tabular}
\end{table}

\subsection{Data Efficiency}

To assess the effect of physics embedding on data efficiency, we trained three multi-channel variants: a purely data-driven model without an explicit reconstruction operator (Identity), and two operator-embedded models incorporating differentiable DAS and Stolt $f$-$k$ migration. Each model was trained using $64$, $128$, $256$, $512$, and $1024$ samples. As shown in Fig.~\ref{fig:data_size}, the operator-embedded models are particularly effective in the very small-data regime, maintaining good performance with only $64$ training samples. In contrast, the purely data-driven model requires substantially more data to achieve comparable performance. These results indicate that incorporating prior knowledge of the reconstruction process reduces the dependence on large training datasets.

\subsection{Transmission generalization}

\begin{figure}[tbp]
    \centering
    \includegraphics[width=\linewidth]{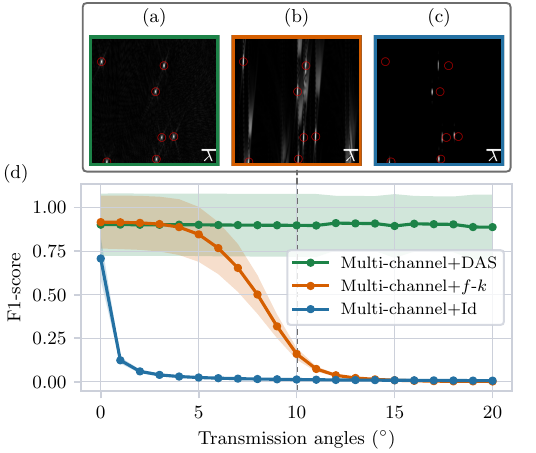}
    \caption{Transmission-angle generalization for steered PW acquisitions. Models were trained only at \(0^\circ\) and evaluated over increasing unseen angles. At \(10^\circ\): (a) Multi-channel+DAS maintains a consistent reconstruction, (b) Multi-channel+\(f\)-\(k\) exhibits ghost artifacts, and (c) Multi-channel+Id produces a spatially shifted reconstruction. (d) \(F_1\) score versus PW steering angle. The solid lines and shaded regions denote the mean and standard deviation, respectively.}
    \label{fig:angles}
\end{figure}

\subsubsection{PW Steering angles}

Transmit steering is central to ultrafast PW imaging, where angular compounding improves image quality while preserving a high frame rate, but it also increases learning complexity. A fully data-driven model requires diverse and balanced training angles to learn angle-dependent RF variations and generalize to unseen configurations, increasing simulation and training demands. In the beamformer-constrained models, the steering angle is incorporated only at inference when constructing the beamforming tensors. This avoids angle-specific training because the MB signatures retain their parabolic geometry across transmit wavefronts.

Fig.~\ref{fig:angles} evaluates inference angles up to \(20^\circ\). Multi-channel+Id degrades sharply when the angle differs from that used during training. This sensitivity is amplified by the strict localization tolerance of approximately \(\lambda/25\), for which small angular mismatches can displace responses outside the accepted region. Multi-channel+DAS remains stable across steering angles, whereas Multi-channel+\(f\)-\(k\) progressively degrades as coherent ghost artifacts appear along the steering direction, increasing FPs despite correct localization of several main MB responses. Additional distortion of the MB responses becomes more pronounced at larger steering angles, reflecting limitations of the underlying migration formulation.

This degradation may reflect a mismatch between the learned RF representation and the global spectral-consistency assumptions of Stolt's \(f\)-\(k\) migration. Unlike the local delay-based sampling of DAS, Stolt's method treats the complete RF dataset as a coherent wavefield and applies steering-dependent phase compensation followed by spectral remapping~\cite{6587395}. Nonphysical phase or spectral components learned under zero-angle supervision may therefore emerge after nonzero-angle compensation and interpolation, producing the observed ghost artifacts.

% spectral corrections -> global not local like DAS (fetch what we want from the parabola) -> Lu also have the same problem -> Fourier-based problem

\subsubsection{DW transmission}

\begin{figure}[tbp]
\centering
\includegraphics[width=\linewidth]{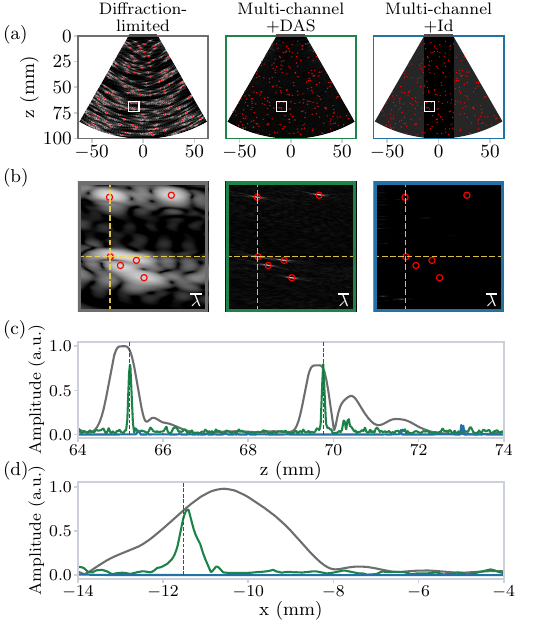}
   \caption{Diverging-wave reconstruction of a simulated frame containing 250 MBs. The columns show the diffraction-limited, Multi-channel+DAS, and Multi-channel+Id reconstructions. (a) Full-sector views with GT bubble locations marked by red circles and the selected ROI outlined by white boxes. The dark region outside the native Cartesian support of the Multi-channel+Id output corresponds to areas without predictions. (b) Zoomed-in views of the selected ROI, with yellow dashed vertical and horizontal lines indicating the axial and lateral cross-sections, respectively. (c) Axial and (d) lateral amplitude profiles extracted along these lines, normalized by each reconstruction's maximum amplitude within the ROI. Red dashed lines indicate the corresponding GT bubble depths in (c) and lateral positions in (d).}

    \label{fig:DW}
\end{figure}

To evaluate generalization beyond PW steering, we applied the Multi-channel+DAS trained exclusively on unsteered PW data to a DW acquisition without retraining. The DAS reconstruction delays were adapted to the DW transmit geometry at inference. Fig.~\ref{fig:DW} presents a simulated frame containing 250 MBs.

Multi-channel+DAS produces compact responses near the GT locations across the displayed sector. In the selected ROI, MB responses that overlap in the diffraction-limited image are more clearly separated, and the axial profile exhibits narrow peaks near the marked GT depths. Multi-channel+Id does not recover the corresponding peaks, even though the selected ROI lies within its native Cartesian prediction domain. Outside that domain, the Multi-channel+Id model provides no predictions, as indicated by the shaded regions in the full-sector view.

\subsection{High Microbubble Density}

\begin{figure}[tbp]
    \centering
    \includegraphics[width=\linewidth]{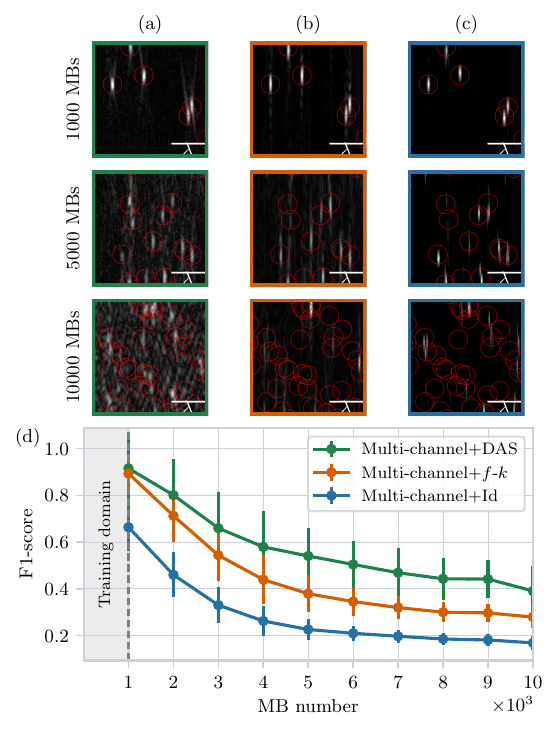}
    \caption{$F_1$ score performance for microbubble densities outside the training domain. (a) Results for the Multi-channel+DAS model at $10^3$, $5\times10^3$, and $10^4$ MBs. (b) Corresponding results for the Multi-channel+$f$-$k$ model. (c) Corresponding results for the Multi-channel+Id model. (d) $F_1$ score as a function of the number of MBs for the different models. Markers indicate the mean $F_1$ score, and error bars indicate one standard deviation.}
    \label{fig:density_adaptive-sat}
\end{figure}

The models were trained on populations of \(10\)–\(1000\) MBs. Robustness beyond this range was evaluated using \(100\) new samples at each fixed population from \(10^3\) to \(10^4\) MBs in increments of \(10^3\), as shown in Fig.~\ref{fig:density_adaptive-sat}. Except for the \(10^3\)-MB boundary case, all evaluated densities were outside the training domain. Overall, the physics-constrained models outperformed the fully data-driven model under high-density conditions.

Multi-channel+Id maintains a high SNR, but its recall decreases markedly with MB density, suggesting that it reconstructs only a subset of the responses while suppressing others, possibly because of strong overlap between RF parabolas. This behavior is also visible in Fig.~\ref{fig:qualitative_res}, where some RF structures are enhanced while others are attenuated.

Multi-channel+\(f\)-\(k\) is more robust at high densities, maintaining precision comparable to Multi-channel+Id while achieving higher recall as the MB population increases. In contrast, Multi-channel+DAS produces stronger background responses, reducing precision when residual intensities exceed the \(1/e\) threshold and are counted as FPs. However, its higher recall indicates detection of a larger fraction of the true MB population. Thus, the DAS-constrained model favors sensitivity, whereas the Id-based model favors compact responses and background suppression.

\section{Discussion}
\label{sec:discussion}

This work introduces a physics-constrained multi-channel RF deconvolution framework for MB ultrasound super-resolution. Unlike the single-channel approach of Blanken et al.~\cite{9755198}, which processes each RF line independently, the proposed anisotropic kernels couple neighboring receive elements and exploit spatial coherence across channels. Learning these interactions before reconstruction improves the separation of overlapping MB responses, as supported by Fig.~\ref{fig:qualitative_res} and Table~\ref{tab:metrics}. In the representative comparison, the multi-channel extension increased \(F_1\) from \(0.90\) for Single-channel+DAS to \(0.92\) for Multi-channel+DAS, alongside improved lateral separation.

Embedding a known beamforming operator further improves detection by decomposing the RF-to-image mapping into learnable RF deconvolution followed by physical reconstruction, as defined in~\eqref{eq:non-linear-mapping}. DAS also introduces implicit inter-channel coupling through coherent summation and image-domain supervision. Even when the deconvolution network processes each RF channel independently, the channel outputs are jointly constrained through the reconstructed image. This constrains the solution space and reduces learning complexity. The benefit is particularly evident in Fig.~\ref{fig:data_size}, where the physics-constrained models perform well with as few as 64 training samples, whereas the fully data-driven model must learn both the MB representation and image formation. Similar benefits from incorporating image-formation physics have been reported in MRI and CT~\cite{Baguer_2020,mrm.28378}. This data efficiency is important because producing large simulated or experimental RF datasets is computationally demanding and time-consuming.

The beamformer also shapes the learned RF representation. DAS performs local delay-based coherent summation~\cite{PERROT2021106309}, whereas Stolt's \(f\)-\(k\) migration applies global spectral remapping and Fourier-domain angle compensation~\cite{6587395}. Accordingly, the DAS-constrained model learns a broader tripolar response along the expected delay trajectory, whereas the \(f\)-\(k\)-constrained model concentrates the response near the echo apex (Figs.~\ref{fig:qualitative_res} and~\ref{fig:rf}). The reconstruction layer therefore structures the intermediate RF representation in a beamformer-dependent manner rather than merely supervising the final image.

Detection performance, localization error, and uncertainty must be interpreted jointly. Compact responses do not necessarily imply complete detection, and higher recall can coexist with stronger residual background. REM aims to complement metrics computed from extracted coordinates~\cite{9130724,9062600,9715054,10497610} and image-level structural measures such as Dice and IoU~\cite{9062600} by analyzing the local predicted response around each GT MB. However, because localization error and uncertainty are computed only for TPs, a model detecting only the easiest MBs may appear artificially accurate. REM may also become ambiguous at extremely high densities because neighboring responses overlap. The resulting scores additionally depend on the intensity threshold, spatial tolerance, component-size criterion, and saturation procedure. These metrics should therefore always be interpreted alongside \(F_1\).

The generalization experiments highlight differences between the embedded operators under changes in transmission. Across steering angles, the DAS-based model remains stable, whereas the \(f\)-\(k\)-based model exhibits coherent ghosts along the steering direction and increasing distortion of MB responses at larger angles (Fig.~\ref{fig:angles}). Residual spectral inconsistencies in the RF representation learned at zero angle could interact with steering-dependent compensation and interpolation, contributing to the observed ghosts. Angle-dependent limitations of the migration formulation may also contribute to response distortion; these possible causes were not isolated in the present study The DW experiment (Fig.~\ref{fig:DW}) extends this to a change from planar to curved wavefronts. Adapting the DAS delays and reconstruction grid at inference preserved localized super-resolved responses without retraining the RF deconvolution network. This also enabled DW reconstruction in polar coordinates, demonstrating flexibility in both transmit geometry and the output grid In contrast, Multi-channel+Id misses responses within the same ROI, indicating stronger dependence on transmission configurations represented during training and suggesting that broader training conditions may be required for comparable generalization. Together, these results support transfer of the DAS-based model across the tested transmission geometries and reconstruction grids without retraining the deconvolution network. The \(f\)-\(k\) approach was not evaluated for DW because its current formulation is restricted to PW imaging.

The density shift in Fig.~\ref{fig:density_adaptive-sat} probes a different form of robustness, namely increasing MB overlap. The physics-constrained models showed modest detection gains over Multi-channel+Id outside the training density range with DAS favoring sensitivity at the cost of stronger background responses, Multi-channel+Id favoring suppression but missing more MBs, and \(f\)-\(k\) providing stronger suppression at high densities. The preferred operator therefore depends on the desired sensitivity-suppression trade-off.

The models were trained on monodisperse MB populations, \(R_0\sim\mathcal{N}\left(2.4~\mu\mathrm{m},(0.14~\mu\mathrm{m})^2\right)\), whose consistent nonlinear signatures reduce acoustic variability and may facilitate learning~\cite{SEGERS20181482,OVERVELDE20102080}. Since the intended application uses monodisperse contrast agents, this reduced variability is an advantage of the selected MB population.

Several limitations remain. The study relies on simulated RF data and a simplified pressure distribution \(P(z,t)\), rather than a realistic three-dimensional field \(P(x,y,z,t)\). The simulations also assumed spatially uniform sound speed and attenuation properties. The effects of spatial variations in these medium properties on RF deconvolution and beamforming remain to be evaluated. Experimental CEUS validation is therefore needed to assess robustness to clutter, motion, tissue background, and acquisition variability, while frameworks such as PROTEUS~\cite{10597664} could provide more realistic simulations. Future work should also investigate training under extremely dense microbubble conditions.

\section{Conclusion}
\label{sec:conclusion}

We proposed a physics-based deep learning framework combining multi-channel RF deconvolution with differentiable beamforming for microbubble ultrasound super-resolution. The framework improved detection, lateral resolution, and data efficiency. Multi-channel+DAS achieved an \(F_1\) score of \(0.92\) and demonstrated generalization across transmit geometries, including the tested steered plane-wave and diverging-wave transmissions, without retraining. Adapting the beamforming delays and reconstruction grid at inference also enabled polar-coordinate reconstruction for diverging-wave imaging. These findings support differentiable beamforming as a constraint for data-efficient super-resolution with flexible transmit geometries and reconstruction grids.

\section*{Acknowledgements}
This work was supported by the European Research Council (ERC) under the European Union's Horizon Europe Research and Innovation Programme through the Super-FALCON Project under Grant 101076844 and by KWF, TKI-Life Sciences and Health and Seno Medical Instruments in project THYNAS+.

% Compact references, keeping each entry together.
\bibliographystyle{unsrtnat}
\begingroup
\interlinepenalty=10000
\footnotesize
\setlength{\bibsep}{3pt}
\bibliography{references}
\endgroup

\end{document}